\documentclass[preprint,aps,showpacs,showkeys,tightenlines,superscriptaddress,fleqn,floatfix]{revtex4}
\usepackage[dvips]{graphicx}
\usepackage{amsmath}
\usepackage[dvips]{graphicx}
\newcommand{\vi}[1]{\mbox{\boldmath $#1$}}

\newcommand{\m}{\thickspace\thickspace\medspace}
\begin{document}
\title{Removal of forbidden states in a three-$\alpha$ system}
\author{H. Matsumura}
\email{hideki@nt.sc.niigata-u.ac.jp}
\affiliation{Graduate School of Science and Technology, Niigata
University, Niigata 950-2181, Japan}
\author{M. Orabi}
\email{orabi@nt.sc.niigata-u.ac.jp}
\affiliation{Graduate School of Science and Technology, Niigata
University, Niigata 950-2181, Japan}
\author{Y. Suzuki}
\email{suzuki@nt.sc.niigata-u.ac.jp}
\affiliation{Department of Physics, and Graduate School of Science and 
Technology, Niigata University, Niigata 950-2181, Japan}
\author{Y. Fujiwara}
\email{fujiwara@ruby.scphys.kyoto-u.ac.jp}
\affiliation{Department of Physics, Kyoto University, Kyoto 606-8502, Japan}
\begin{abstract}
The ground and excited 0$^+$ states of $^{12}$C are investigated in a 
3$\alpha$ macroscopic model using the deep potential of Buck, Friedrich 
and Wheatley. The elimination of forbidden states is performed either 
by constructing the allowed state space explicitly or by using the 
orthogonalizing pseudopotential. The well-known enigmatic behavior 
of the latter approach is resolved. It is safe to define the 
forbidden states referring to the underlying microscopic model. 
\end{abstract}
\maketitle

\section{Introduction}

A microscopic multicluster model is successful for understanding  
the structure of light nuclei~\cite{mmc}. One example recently applied to halo 
nuclei is the description of lithium 
isotopes $^{7-11}$Li in terms of an $\alpha + t+xn$ 
model~\cite{Li.isotope} including $(2+x)$ clusters, 
where $x$ denotes the number of neutrons, i.e., $x=0,1,\ldots, 4$ for 
$^{7-11}$Li. With the use of an effective two-nucleon potential, this
approach reproduces consistently the ground state 
energies of these nuclei together with their proton and neutron 
radii. In particular, a recent experiment~\cite{Li11radius} 
has confirmed that the isotope dependence 
of the charge radii is best reproduced by the 
stochastic variational calculation~\cite{book} on the basis of this 
multicluster model. 

In the microscopic cluster model, the structures of clusters are described 
in terms of appropriate wave functions, usually in the simplest
configurations of the shell model, and an antisymmetry 
requirement on the total wave function is taken into account exactly. 
The multicluster model is thus 
a microscopic many-body theory based on 
an effective nucleon-nucleon potential. One must, however, 
compute fairly complicated matrix elements for those non-orthogonal 
basis functions used in the multicluster model, which makes it hard 
to apply to larger nuclei. This motivates another treatment of 
the multicluster system, that is, a macroscopic cluster model. 

In a macroscopic cluster model, the clusters are treated as 
structureless point particles. In this case, cluster-cluster 
interactions are usually represented by appropriate local
potentials which are, more or less, phenomenologically determined 
so as to reproduce the relevant data on the cluster-cluster systems. 
One of the simplest and well-investigated examples is a 3$\alpha$ model
for $^{12}$C. 

Many calculations based on the microscopic 3$\alpha$ model show that 
it reproduces the general features of the low-lying spectrum and the 
spectroscopic properties of $^{12}$C~\cite{kamimura,uegaki}. Thus we would 
expect that the macroscopic version of the 3$\alpha$ system could give 
results of similar quality if it would be a realistic substitute of 
the microscopic model. Since the $\alpha$ particle is a tightly bound system, 
it is natural to conceive that this expectation may come true. Two kinds of local 
$\alpha\!-\!\alpha$ potential are often employed in the macroscopic 
3$\alpha$ model. 
One is an angular momentum dependent potential such as 
Ali-Bodmer (AB) potential~\cite{ali}, or that proposed in
Ref.~\cite{chien}. These potentials are repulsive at the short 
$\alpha\!-\!\alpha$ distance in the $s$ and $d$ waves.   
The other is a deep, angular momentum independent potential 
such as Buck-Friedrich-Wheatley (BFW) potential~\cite{bfw,friedrich}. 
Both potentials reproduce 
the phase shifts of low-energy $\alpha\!-\!\alpha$ elastic 
scatterings well, but 
they differ in the prediction about a 2$\alpha$ bound state. The AB potential 
produces no bound state, consistently 
with the fact that the 2$\alpha$ system has no bound states. 
Contrary to this, the BFW potential is so deep that it produces 
two bound states ($0s$ and $1s$) in the $s$ wave and one bound 
state ($0d$) in the $d$ wave. These 
bound states are considered redundant or forbidden states from which a  
physically acceptable $\alpha\!-\!\alpha$ relative motion function must 
be free. The removal of unphysical bound states has been done using 
either the projection technique with an orthogonalizing 
pseudopotential $\lambda \Gamma$~\cite{kukulin} or 
a supersymmetric transform of the original potential~\cite{baye}. 
It was shown in Ref.~\cite{baye} that the supersymmetric partner 
potential of the BFW potential is similar to the AB potential. 
 
How well does the macroscopic model with the AB or BFW potential 
reproduce the results of the
underlying microscopic calculation for $^{12}$C? In this respect, the macroscopic 
calculation gives quite poor results~\cite{tursunov,takahashi,descouvemont,tursunov2}, 
e.g., the ground state energy appears just 
below the 3$\alpha$ threshold, which is far from the experimental 
value, $-7.27$ MeV. Moreover, the calculated ground state hardly has the properties 
suitable for the $^{12}$C ground state. It is also 
confirmed that the projection technique gives highly unstable results 
which depend on the projection constant 
$\lambda$ and on how accurately the forbidden
states are solved as well~\cite{tursunov,descouvemont}. 
A convergent result is obtained only by letting $\lambda$ sufficiently large.  
These results are quite embarrassing considering the success of the 
microscopic 3$\alpha$ model. 
 
The purpose of the present study is to clarify the origin of the
enigmatic behavior of the projection technique as well as the reason why its 
ground state obtained in the macroscopic 3$\alpha$ model is far from 
the physical ground state. To answer these questions, 
we carry out a direct method of removing the forbidden states by diagonalizing the projection operator $\Gamma$. We will show, using an example 
of the BFW potential, that 
the enigma can be understood by the existence of 
particular 3$\alpha$ states, which are excluded from the 3$\alpha$ 
function space in the projection technique, but, in our opinion, 
should be included as allowed states. 
This is demonstrated by defining both of allowed and forbidden states 
in translationally invariant harmonic-oscillator basis functions. 

In the microscopic 3$\alpha$ model, the $\alpha$ particle is usually 
described in terms of a harmonic-oscillator $(0s)^4$ configuration. 
Then it is natural to define allowed or forbidden states referring to 
the Pauli exclusion principle. 
Such allowed states are obtained explicitly for the 3$\alpha$ system 
through the norm kernel of the microscopic 3$\alpha$ wave  
function~\cite{matsumura}. A simpler approximation is to remove the 2$\alpha$ 
Pauli-forbidden states, namely the $0s, \, 1s$ and  $0d$ harmonic-oscillator 
states, from any pair of the $\alpha-\alpha$ relative 
motion~\cite{smirnov,horiuchi}. We call the resulting states  
``Pauli-allowed'' states, and distinguish them from 
those defined according to the redundant bound states in the BFW 
potential. 

In sect.~\ref{removal}, we discuss the methods of removing the forbidden 
states in the 3$\alpha$ system. We stress the utility of the
harmonic-oscillator basis functions, which makes it straightforward 
to construct the allowed states. In sect.~\ref{ashosm}, we discuss the 
Pauli-allowed states which have direct relationship to 
the microscopic 3$\alpha$ model, where 
the $\alpha$ particle is described in terms of a harmonic-oscillator 
$(0s)^4$ configuration. 
In sect.~\ref{bfwcase}, we show how the enigmatic behavior of the 
projection technique can be understood. In sect.~\ref{alpha-alphapot}, 
we discuss such an $\alpha$-$\alpha$ potential that is  
suited for the macroscopic calculations. A summary is 
given in sect.~\ref{summary}. Some of the lowest 
Pauli-allowed states are given in Appendix.

\section{Solution with the removal of forbidden states}
\label{removal}

The Hamiltonian for the 3$\alpha$ system is in the form
\begin{equation}
H=\sum_{i=1}^3\frac{{\vi P}_i^2}{2m_{\alpha}}-T_{\rm c.m.}+\sum_{i=1}^3V_{i},
\end{equation}
where $m_{\alpha}$ is the mass of the $\alpha$ particle and the kinetic energy of 
the center of mass motion is subtracted.  
$V_{i}$ is the $\alpha\!-\!\alpha$ potential acting between $(jk)$ pair.  Here we use the 
convention of $(jk)=(23)$ for $i=1$, $(jk)=(31)$ for $i=2$, and 
$(jk)=(12)$ for $i=3$. We want to solve 
\begin{equation}
H\Psi=E\Psi,
\label{sch-eq}
\end{equation}
with the constraint 
\begin{equation}
\langle \varphi_{f}({\vi r}_i) \mid \Psi \rangle=0   \ \ \ (i=1,2,3),
\label{condition}
\end{equation}
where $\varphi_f$ stands for a particular forbidden state. 
When we have several 
forbidden states, the above condition must be applied to all of them. The $\alpha\!-\!\alpha$ 
relative distance vector is given by ${\vi r}_i\!=\!{\vi R}_j-{\vi R}_k$, where 
${\vi R}_i$ is the center of mass coordinate of the $i$th $\alpha$ particle. 
Another relative coordinate is denoted by ${\vi \rho}_i\!=\!({\vi R}_j+{\vi R}_k)/2-{\vi R}_i$. 
The wave function $\Psi$ must be totally symmetric with respect to the interchange 
of ${\vi R}_i$'s.
 
As in the BFW potential, we assume that there exist three forbidden states, 
0$s$, 1$s$ and 0$d$. The function $\varphi_f$ in 
Eq.~(\ref{condition}) thus represents these three states. 
To solve Eq.~(\ref{sch-eq}) with the condition~(\ref{condition}), we 
define an operator 

\begin{equation}
\Gamma=\sum_{i=1}^3\Gamma_{i},
\label{def.gamma} 
\end{equation}
where
\begin{equation}
\Gamma_i=\Gamma_i(0s)+\Gamma_i(1s)+\Gamma_i(0d)
\end{equation}
with
\begin{equation}
\Gamma_i(n\ell)=\sum_{m=-\ell}^{\ell} \mid \varphi_{n\ell m}({\vi r}_i)\rangle 
\langle  \varphi_{n\ell m}({\vi r}_i)\mid.
\label{bfwproj}
\end{equation}
The operator $\Gamma_i$ is a projector onto the forbidden states for 
the relative motion between the $\alpha$ particles of $(jk)$ pair, 
but we must note that, for $i\ne j$, it satisfies the relation  
$\Gamma_i \Gamma_j \ne \Gamma_j \Gamma_i\ne 0$.  

We consider the eigenvalue problem of $\Gamma$
\begin{equation}
\Gamma \, \chi_{LM}^{[3]\gamma}=\gamma \, \chi_{LM}^{[3]\gamma},
\label{eig.gamma}
\end{equation}
and obtain those eigenfunctions which are totally symmetric, i.e., [3] symmetry. 
The eigenvalue $\gamma$ is non-negative. This is seen as follows:
\begin{eqnarray}
\gamma&=&\langle \chi_{LM}^{[3]\gamma}\mid \Gamma \mid \chi_{LM}^{[3]\gamma} \rangle
\nonumber \\
&=&3\, \langle \chi_{LM}^{[3]\gamma}\mid \Gamma_i \mid \chi_{LM}^{[3]\gamma} \rangle
\ \ \ \ \ (i=1,2,3) \nonumber \\
&=&3\sum_f \langle \chi_{LM}^{[3]\gamma}\mid \varphi_f({\vi r}_i) \rangle 
\langle \varphi_f({\vi r}_i) \mid \chi_{LM}^{[3]\gamma} \rangle. 
\end{eqnarray}
Here we used the symmetry of $\chi_{LM}^{[3]\gamma}$ and noted that 
the matrix element in the last line is non-negative, which follows from the fact that 
$\langle \varphi_f({\vi r}_i) \mid \chi_{LM}^{[3]\gamma} \rangle$ is a function of 
${\vi \rho}_i$, say, $v({\vi \rho}_i)$, and thus $\langle \chi_{LM}^{[3]\gamma}\mid \varphi_f({\vi r}_i) \rangle 
\langle \varphi_f({\vi r}_i) \mid \chi_{LM}^{[3]\gamma} \rangle=\langle v \mid v 
\rangle$ can not be negative. Clearly $\gamma=0$ is equivalent to   
$v=\langle \varphi_f({\vi r}_i) \mid \chi_{LM}^{[3]\gamma} \rangle=0 \, (i=1,2,3)$, 
namely, the eigenfunction with $\gamma=0$ contains no forbidden components, so 
it satisfies the condition~(\ref{condition}). We can thus expand the solution of 
Eq.~(\ref{sch-eq}) in terms of the eigenfunctions belonging to $\gamma=0$, 
\begin{equation}
\Psi=\sum_p C_p \chi_{pLM}^{[3]\gamma=0},
\label{sol.red-free}
\end{equation}
where $p$ is introduced to distinguish the orthonormal eigenfunctions with 
$\gamma=0$. The coefficients $C_p$ and the energy $E$ 
are determined from the secular equation
\begin{equation}
\sum_{p'} \langle \chi_{pLM}^{[3]\gamma=0} \mid H-E \mid \chi_{p'LM}^{[3]\gamma=0}
 \rangle C_{p'}=0.
\label{sol1}
\end{equation}
The procedure explained above is a standard way to obtain a solution
which is free from any forbidden components. One, however, has to note
that to obtain vanishing eigenvalues of Eq.~(\ref{eig.gamma}) 
demands a calculation of high accuracy. 

Another method of solution equivalent to the above direct method is to add 
the orthogonalizing pseudopotential~\cite{kukulin} to the Hamiltonian 
\begin{equation}
\widetilde{H}=H+\lambda \Gamma
\end{equation}
and solve the equation for $\Psi(\lambda)$ of [3] symmetry 
\begin{equation}
\widetilde{H}\Psi(\lambda) =E(\lambda) \Psi(\lambda)
\end{equation}
as a function of the projection constant $\lambda$. The solution for 
$\Psi(\lambda)$ may be expanded as in Eq.~(\ref{sol.red-free}) but then 
all the eigenfunctions with both $\gamma =0$ and $\gamma \ne 0$ must 
be included. The desired solution is such that   
$E(\lambda)$ and $\Psi(\lambda)$ are stable for sufficiently 
large $\lambda$. To understand this, we note that the energy 
$E(\lambda)$ reads as follows:
\begin{equation}
E(\lambda)=\langle \Psi(\lambda) \mid H \mid \Psi(\lambda) \rangle 
+ \lambda \langle \Psi(\lambda) \mid \Gamma \mid \Psi(\lambda) \rangle.
\label{e.lambda}
\end{equation} 
The first term on the right-hand side of  Eq.~(\ref{e.lambda}) is expected to 
change moderately as a function of $\lambda$, but the second term  increases 
rapidly with increasing $\lambda$ in such a way that $\Psi(\lambda)$ has 
a significant overlap with the forbidden states because in that case 
the probability of finding the forbidden components in 
$\Psi(\lambda)$, $\langle \Psi(\lambda) \mid 
\Gamma \mid \Psi(\lambda) \rangle$, becomes a positive, non-vanishing value. 
In contrast to 
this, $E(\lambda)$ can be stable for sufficiently large $\lambda$, 
provided that $\Psi(\lambda)$ has vanishingly small or no overlap with 
the forbidden states. 
The advantage of this pseudopotential method is that one does not need to 
solve Eq.~(\ref{eig.gamma}) to construct the allowed space spanned by 
the solutions with $\gamma=0$, but only needs 
to diagonalize the Hamiltonian with the pseudopotential. To obtain a 
stable solution, however, one has to provide a function space large 
enough to remove the forbidden states. 

It is important for our purpose to construct a complete set of [3] 
symmetry for the 
3$\alpha$ system. For example, the eigenvalue problem (\ref{eig.gamma}) of 
the operator $\Gamma$ can be solved in such a set. 
The construction of the complete set can be performed using 
translationally invariant harmonic-oscillator functions 
\begin{equation}
[\psi_{\ell_1}^{q_1}({\vi r}_1)\psi_{\ell_2}^{q_2}({\vi \rho}_1)]_{LM},
\label{hobasis}
\end{equation}
where $\psi_{\ell m}^{q}$ is 
the oscillator function with the number of oscillator 
quanta $q$, the orbital angular momentum $\ell$ and its $z$ 
component $m$. ($q\!-\!\ell$ must be a non-negative even number.) 
The symbol $[\cdots]_{LM}$ indicates the angular momentum 
coupling. The oscillator parameters of 
the functions $\psi_{\ell_1m_1}^{q_1}({\vi r}_1)$ and 
$\psi_{\ell_2m_2}^{q_2}({\vi \rho}_1)$ 
are $2\nu$ and $\frac{8}{3}\nu$, respectively. Here 
$\nu=(m_{N}\omega /2\hbar)$ is the 
oscillator parameter of the single-nucleon wave function in the harmonic 
oscillator shell model, where $m_N$ is the nucleon mass. 
We choose $\nu=0.26$ fm$^{-2}$ because it is an appropriate value to 
reproduce the charge radius of the $\alpha$ particle using the $(0s)^4$ 
wave function~\cite{matsumura}. One may choose other values of $\nu$, but then 
the diagonalization of the operator $\Gamma$ and the Hamiltonian,  
Eqs.~(\ref{eig.gamma}) and (\ref{sol1}), would require larger dimension. 
Though an 
elegant method of constructing the states with definite permutational 
symmetry is given in Ref.~\cite{moshinsky}, we follow a simple 
procedure of diagonalizing the symmetrizer 
\begin{equation}
{\cal S}=\frac{1}{3}(1+(1,2,3)+(1,3,2)),
\end{equation}
where $(1,2,3)$ and $(1,3,2)$ are cyclic permutations. The states with 
[3] symmetry can be obtained as those with unit eigenvalue of ${\cal S}$. 
The symmetrizer ${\cal S}$ conserves the number of total oscillator quanta 
$Q=q_1+q_2$, so the diagonalization of ${\cal S}$ can be done 
separately for each $Q$.  
In this diagonalization one needs the matrix element of the permutation 
$P$ ((1,2,3) and (1,3,2)) 
in the basis~(\ref{hobasis}), which can be easily evaluated using 
a generalized Talmi-Moshinsky bracket~\cite{moshinsky,suzuki74}.

\section{Pauli-allowed states in harmonic-oscillator basis}
\label{ashosm}

So far we have assumed that the forbidden states $\varphi_f({\vi
r})$ are the bound states of the $\alpha\!-\!\alpha$ potential, which 
is considered natural in several 
studies~\cite{smirnov,kukulin,tursunov,descouvemont,tursunov2}. 
A different viewpoint is that the forbidden or redundant states should be
defined according to the underlying microscopic model, that is, they 
are determined solely from the wave function of the $\alpha$ particle, not from the $\alpha\!-\!\alpha$ potential employed. For the 
$\alpha$ particle described with the $(0s)^4$ harmonic-oscillator 
shell-model with the oscillator constant $\nu$, 
it is well-known that three Pauli-forbidden states appear for the $\alpha\!-\!\alpha$ 
relative motion. They have 
the same quantum numbers, $0s, 1s$ and $0d$, as those of the BFW potential, 
but their wave functions are simply given by the harmonic-oscillator functions 
$\psi_{\ell m}^{q} (q\!=\!0, 2)$. This approach is in fact a 
multicluster version of the orthogonality condition 
model~\cite{horiuchi,ocmhori}. An $\alpha\!-\!\alpha$ potential 
used in this case is often taken to be close to a double 
folding potential~\cite{kurokawa}. 

It is useful to define an operator $\Gamma^{\rm HO}$, 
an analogue to the operator (\ref{def.gamma}), 
in the harmonic-oscillator basis, where $\Gamma_i(n\ell)$ of 
Eq.~(\ref{bfwproj}) is replaced by 
\begin{equation}
\Gamma_i^{\rm HO}(n\ell)=\sum_{m=-\ell}^{\ell} \mid \psi_{\ell m}^{q}({\vi r}_i)\rangle 
\langle  \psi_{\ell m}^{q}({\vi r}_i)\mid\ \ \ (q=2n+\ell).
\label{hoproj}
\end{equation}
The eigenvalue problem for $\Gamma^{\rm HO}$, the analogue of 
Eq.~(\ref{eig.gamma}), can be solved in the basis (\ref{hobasis}). 
The eigenfunctions can be classified by $Q$, an SU(3) irreducible 
representation $(\lambda \mu)$ and an additional quantum number 
$\kappa$~\cite{smirnov,horiuchi} ($\kappa$ serves to distinguish a multiple occurrence of 
the same $(\lambda \mu)$ states, and it is otherwise suppressed). That is, 
\begin{equation}
\chi_{LM}^{[3]Q(\lambda \mu)\kappa \gamma},
\end{equation}
and the eigenvalues $\gamma$ are either zero (Pauli-allowed) or 
non-zero values of order unity (Pauli-forbidden). 
The SU(3) coupled basis is easily constructed by use of SU(3) 
coefficients~\cite{draayer}:
\begin{equation}
[\psi^{q_1}({\vi r})\psi^{q_2}({\vi \rho})]^{(\lambda \mu)}_{LM}=\sum_{\ell_1 \ell_2}
\langle (q_10)\ell_1, (q_20)\ell_2 \mid\mid(\lambda \mu) L\rangle \, 
[\psi^{q_1}_{\ell_1}({\vi r})\psi^{q_2}_{\ell_2}({\vi \rho})]_{LM}.
\label{su3basis}
\end{equation}
Here $(\lambda \mu)$ can in general take the representations of 
$(q_1\!+\!q_2,0), (q_1\!+\!q_2\!-\!2,1), (q_1\!+\!q_2\!-\!4,2), \ldots, 
(q_>\!-\!q_<,q_<)$, where 
$q_>\!=\!{\rm max}(q_1,q_2)$ and $q_<\!=\!{\rm min}(q_1,q_2)$, but is limited 
to those which contain the angular momentum $L$. For $L=0$, both 
$\lambda$ and $\mu$ must be even, and a concise expression for the SU(3) 
coefficients with $L=0$ is given in Ref.~\cite{hecht}. 

For $Q < 8$, all the [3]-symmetry basis 
states are Pauli-forbidden. For $Q=8$ we have only one Pauli-allowed
state with $(\lambda \mu)\!=\!(04)$. 
For $Q\!=\!10$, two Pauli-allowed states, 
$(24)$ and $(62)$, appear. We give in Appendix the Pauli-allowed 
states $\chi_{00}^{[3]Q(\lambda \mu)\kappa \gamma=0}$ for $Q=8-14$  
in the SU(3) coupled basis of Eq.(\ref{su3basis}).

\section{Case of the BFW potential}
\label{bfwcase}

\begin{table}[b]
\caption{Expansion of the three bound states of the BFW potential in terms 
of the harmonic-oscillator functions. See Eq.~(\ref{uexp}). }
\label{bfwspexp}
\begin{center}
\begin{tabular}{lllllll}
\hline \hline 
$n'$ && \quad\quad $U^{00}_{n'}$ && \quad $U^{10}_{n'}$ && \quad $U^{02}_{n'}$    \\
\hline
  0 &&\m 0.991259    \; && 0.122458    \; &&  0.994643      \\
  1 &&$-$0.123881    \; && 0.984707    \; &&  3.8363$\times 10^{-2}$ \\
  2 &&\m 4.4209$\times 10^{-2}$ \; && 3.6841$\times 10^{-2}$ \; &&  9.2784$\times 10^{-2}$  \\
  3 &&$-$9.4402$\times 10^{-3}$ \; && 0.114653    \; &&  1.6585$\times 10^{-2}$   \\
  4 &&\m 3.7620$\times 10^{-3}$ \; && 1.8535$\times 10^{-2}$ \; &&  1.6517$\times 10^{-2}$   \\
  5 &&$-$9.6104$\times 10^{-4}$ \; && 2.0740$\times 10^{-2}$ \; &&  5.7095$\times 10^{-3}$   \\
  6 &&\m 4.2760$\times 10^{-4}$ \; && 6.5918$\times 10^{-3}$ \; &&  4.2427$\times 10^{-3}$   \\
  7 &&$-$1.1796$\times 10^{-4}$ \; && 5.1703$\times 10^{-3}$ \; &&  2.0350$\times 10^{-3}$   \\
  8 &&\m 5.8853$\times 10^{-5}$ \; && 2.3322$\times 10^{-3}$ \; &&  1.3746$\times 10^{-3}$   \\
  9 &&$-$1.6647$\times 10^{-5}$ \; && 1.6078$\times 10^{-3}$ \; &&  7.7920$\times 10^{-4}$   \\
 10 &&\m 9.0250$\times 10^{-6}$ \; && 8.7429$\times 10^{-4}$ \; &&  5.1694$\times 10^{-4}$  \\
 $\vdots$ && \qquad\quad $\vdots$ && \qquad $\vdots$ && \qquad $\vdots$ \\
\hline
\end{tabular}
\end{center}
\end{table}

The BFW potential reproduces very well the $\alpha\!-\!\alpha$ phase shifts of 
the $s, \, d$ and $g$ waves for $E_{\rm c.m.} < 17$ MeV~\cite{bfw}. 
It is a local, angular momentum independent potential given by 
\begin{equation}
V({\vi r})=v_0\exp(-\rho_0 r^2)+\frac{4e^2}{r}{\rm erf}(\beta r),
\end{equation}
where ${\vi r}$ is the $\alpha\!-\!\alpha$ relative distance vector. 
The parameters of the BFW potential are 
$v_0=-122.6225$ MeV, $\rho_0=0.22$ fm$^{-2}$, and $\beta=0.75$ fm$^{-1}$. 
We choose $\frac{\hbar^2}{m_{\alpha}}=10.4465$ MeV$\,$fm$^{2}$ and 
$e^2=1.44$ MeV$\, $fm. The three bound states of the BFW potential, 
$\varphi_{n\ell m}$, are 
expanded in terms of the harmonic-oscillator functions $\psi_{\ell m}^{q'}({\vi r})
\!=\!R_{n'l}(r)Y_{\ell m}(\widehat{\vi r})\ (n'\!=\!(q'\!-\!\ell)/2)$:
\begin{equation}
\varphi_{n\ell m}({\vi r})=\sum_{n'} U_{n'}^{n\ell}R_{n'\ell}(r) Y_{\ell m}(\widehat{\vi r}).
\label{uexp}
\end{equation}
The phase convention of $R_{n'\ell}(r)$ is such that it is positive for 
$r$ greater than the outermost nodal point. 
Table~\ref{bfwspexp} lists the expansion coefficients $U_{n'}^{n\ell}$. It is seen that 
the bound state wave functions $\varphi_{n\ell m}({\vi r})$ are 
well approximated by the oscillator functions 
$\psi_{\ell m}^q({\vi r})\ (q\!=\!2n\!+\!\ell)$, but we will see 
that the small difference between the two functions produces a 
huge effect on the energy of the 3$\alpha$ system.  

\begin{figure}[t]
\begin{center}
\rotatebox{-90}{\includegraphics[width=8cm,clip]{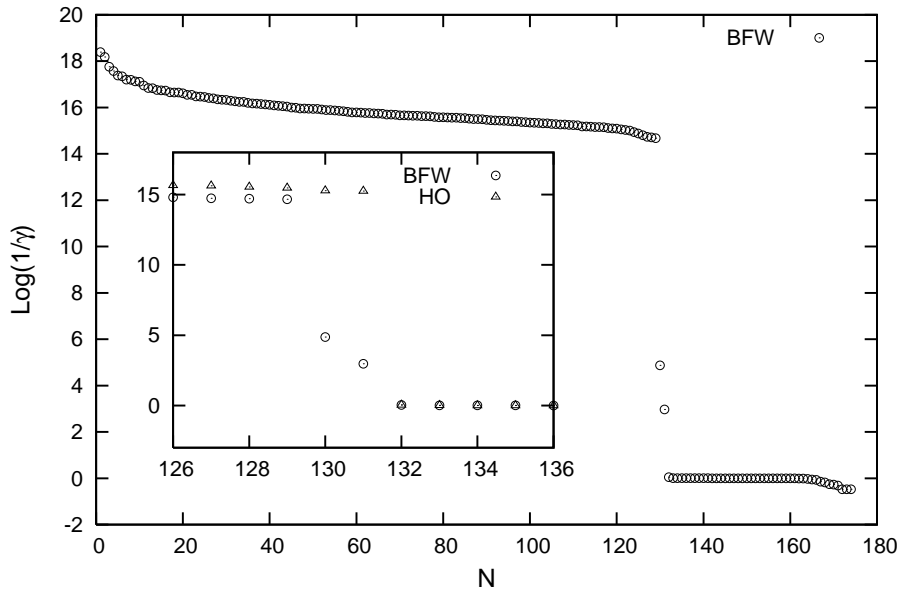}}
\end{center}
\caption{Plot of Log$(1/\gamma)$ for $N=1\!-\!174$ eigenvalues obtained
 for $Q_{\rm max}\!=\!30$, where $\gamma$ is
 the eigenvalue of the operator $\Gamma$ defined by 
 Eq.~(\ref{def.gamma}) using the three forbidden states of the 
 BFW potential (see Eq.~(\ref{eig.gamma})). In the inset, 
 the eigenvalues around $N=130$ are compared with those obtained 
 for $\Gamma^{\rm HO}$. }
\label{inv.gamma}
\end{figure}

The expansion (\ref{uexp}) makes it possible to solve accurately 
the eigenvalue problem (\ref{eig.gamma}) in the translationally invariant 
harmonic-oscillator functions provided the maximum number of total 
oscillator quanta $Q_{\rm max}$ is sufficiently large. 
We treat only $0^+$ case in this paper. 
For $Q_{\rm max}\!=\!30$, there are 444 translationally invariant 
basis functions, and from them 
we can construct 174 independent basis functions of  [3] symmetry. Figure~\ref{inv.gamma} displays Log$(1/\gamma)$ ,where $\gamma$ is the eigenvalue 
of $\Gamma$ obtained for this case. It is clear in the
 figure that the eigenvalues $\gamma$ diagonalized in these [3]-symmetry functions 
are classified into three groups: the first group (Group I) is 
characterized by $\gamma=0$ and it contains 129 eigenfunctions 
(the values of $\gamma$ are actually less than $10^{-14}$, which we 
 considered zero within the present numerical accuracy).  The elements 
of this group are allowed states and constitute the basis for diagonalizing 
the Hamiltonian in Eq.~(\ref{sol1}). The second group (Group II)
contains just two states whose eigenvalues are approximately zero; 
$\gamma_1\!=\!1.35152\times 10^{-5}$ and $\gamma_2\!=\!1.07152\times 
10^{-3}$. The appearance of these states was suggested in a Faddeev treatment of the 
3$\alpha$ system~\cite{fujiwara2,fujiwara3,fujiwara4}. 
The third group (Group III) contains 43 elements with  eigenvalues of order unity ($\sim 1$). 
These last states must be discarded  
because they have a substantial overlap with the forbidden 
states. Differently, using $\Gamma^{\rm HO}$, in which the two-body 
forbidden states are 
the harmonic-oscillator functions as defined in Eq.~(\ref{hoproj}), the eigenvalues are divided into two groups only corresponding to I and III, and there is no such a group like II. The inset of the figure compares the eigenvalues of both cases 
for $N=126-136$. 

\begin{table}[t]
\caption{The number of allowed and forbidden states for the 3$\alpha$ system with 
$0^+$. The $\alpha\!-\!\alpha$ forbidden states are taken as the three bound states of the 
BFW potential. $Q_{\rm max}$ is the maximum number of total oscillator quanta, $N_{\rm basis}$ 
is the basis dimension of the translationally invariant harmonic-oscillator 
functions with $Q=0-Q_{\rm max}$, $N^{[3]}$ is the number of  totally symmetric 
functions with $Q=0-Q_{\rm max}$, $N^{\gamma=0}$ is the number of
 allowed states of 
Group I, $N^{\gamma \approx 0}$ is the number of eigenfunctions 
of Group II, and $N^{\gamma \sim 1}$ is the number of forbidden states 
of Group III. }
\label{Qdep.gamma}
\begin{center}
\begin{tabular}{cccccc}
\hline\hline
$Q_{\rm max}$ & $N_{\rm basis}$ & $N^{[3]}$ & $N^{\gamma=0}$ & $N^{\gamma \approx 0}$ 
& $N^{\gamma \sim 1}$ \\
\hline
30 & 444 & 174 & 129 & 2 & 43 \\
40 & 946  & 358 & 298 & 2 & 58 \\
50 & 1729 & 640 & 565 & 2 & 73 \\
60 & 2856 & 1041 & 951 & 2 & 88 \\
\hline
\end{tabular}
\end{center}
\end{table}

Table~\ref{Qdep.gamma} shows the $Q_{\rm max}$ dependence of the number 
of eigenfunctions belonging to each group. There are two noteworthy
points; first, the two solutions of Group II always appear for all 
$Q_{\rm max}$ values. Moreover, we have confirmed that these solutions 
(the eigenfunctions and the eigenvalues) are very stable
against the increase of $Q_{\rm max}$, e.g., for $Q_{\rm max}\!=\!50$, 
$\gamma_1\!=\!1.35191\times 10^{-5}$ and 
$\gamma_2\!=\!1.07153\times 10^{-3}$.  
Second, the sum of the members of Groups I and II is equal to
the number of Pauli-allowed states which occur for the case of 
$ U_{n'}^{n\ell}=\delta_{n' n}$, namely 
the bound states $\varphi_{n\ell m}$ are replaced with the 
harmonic-oscillator functions $\psi^{2n\!+\!\ell}_{\ell m}$. 

For later discussion we expand the two states of Group II in terms of 
the harmonic-oscillator basis functions. 
Table~\ref{APAS} lists the overlap of these two states with some of the 
Pauli-allowed states $\chi^{[3]Q(\lambda \mu)\kappa \gamma=0}_{00}$ 
(see Appendix for the expressions of these Pauli-allowed states). From the table we can notice that 
the two states with the eigenvalues $\gamma_1$ and $\gamma_2$ 
contain predominantly $Q=10,\, 14$ and $Q=8,\, 12$ components,
respectively.  

\begin{table}[t]
\caption{Expansion coefficients of the two states of Group II in terms
 of the Pauli-allowed states $\chi^{[3]Q(\lambda \mu)\kappa
 \gamma=0}_{00}$. The index $\kappa$ is suppressed when it is 
 unnecessary. See Appendix for the definition of the two independent states 
which appear for $Q(\lambda \mu)=14(64)$.}
\label{APAS}
\begin{center}
\begin{tabular}{lllllll}
\hline\hline
$Q$ && $(\lambda \mu)\kappa$ &&\m $\chi^{[3]\gamma_1}_{00}$ &&\m $\chi^{[3]\gamma_2}_{00}$ \\
\hline
   8 \: &&    (04)    \; &&\m 0.28139 \; &&\m 0.85945      \\
  10 \: &&    (24)    \; &&\m 0.38513 \; &&$-$0.11666      \\
  10 \: &&    (62)    \; &&\m 0.58421 \; &&$-$0.05242      \\
  12 \: &&    (06)    \; &&\m 0.06598 \; &&$-$0.02209      \\
  12 \: &&    (44)    \; &&$-$0.03502 \; &&\m 0.15717      \\
  12 \: &&    (82)    \; &&\m 0.18853 \; &&$-$0.06749      \\
  12 \: &&    (12, 0) \; &&$-$0.12627 \; &&\m 0.42769      \\
  14 \: &&    (26)    \; &&$-$0.07958 \; &&\m 0.01859      \\
  14 \: &&    (64)1   \; &&$-$0.08228 \; &&\m 0.03768      \\
  14 \: &&    (64)2   \; &&\m 0.02034 \; &&\m 0.00586      \\
  14 \: &&    (10, 2) \; &&\m 0.20863 \; &&$-$0.02453      \\
  14 \: &&    (14, 0) \; &&$-$0.42113 \; &&\m 0.11500      \\
\hline
\end{tabular}
\end{center}
\end{table}

In order to understand the reason why the orthogonalizing 
pseudopotential approach 
gives a result quite different from that of the microscopic model, 
we first diagonalize the Hamiltonian $H$ in the following three 
different basis sets and compare their energies and properties for 
the ground and excited 0$^+$ states. The three sets are 
(i) a set spanned by the basis functions of Group I, 
(ii) a set spanned by the basis functions of Groups I and II, 
and (iii) a set spanned by the Pauli-allowed states 
$\chi^{[3]Q(\lambda \mu)\kappa \gamma=0}_{00}$. The calculation has 
been performed for some $Q_{\rm max}$ values to check the convergence 
of the solution. The results obtained 
with the three sets are listed in Table~\ref{comp.0+}. 
The energies of set (i) correspond to the 
eigenvalues of Eq.~(\ref{sol1}), 
and will approach the energies $E(\lambda)$ for $\lambda$ sufficiently 
large as will be shown later. The ground state energy for $Q_{\rm max}=60$ is $-0.22$ MeV, which is consistent with the value obtained 
by the projection technique~\cite{descouvemont,tursunov2}. 
The calculated energy is far from the experimental value $-7.27$ MeV 
and rather close to the 3$\alpha$ threshold where the second $0^+$ state 
is observed. The root mean square (rms) radius $R_{\rm rms}$ 
of the $\alpha$ particle distribution 
for the ground state turns out to be 2.31 fm, 
which is much larger than $1.92$ fm~\cite{matsumura}. 
The latter value is obtained by the 3$\alpha$ boson wave 
function which is mapped from the microscopic 3$\alpha$ wave 
function that reproduces both the energy and charge radius of 
the ground state of $^{12}$C. 
The result of set (ii) is in a sharp 
contrast to case (i). The ground state is now very strongly bound 
and consequently its rms radius is too small compared to the
result of the microscopic calculation, whereas the calculated 
second 0$^+$ state appears near the experimental energy and its 
rms radius is increased to about 2.34 fm, though it is still too small 
compared to 4.26 fm of the microscopic calculation~\cite{matsumura}.  
It is remarkable that the results calculated with 
sets (ii) and (iii) are very similar. From this we may conclude that 
the function space spanned by the members of Groups I and II is nearly  
the same as that of the Pauli-allowed states. 

\begin{table}[t]
\begin{center}
\hspace*{-4cm}
\begin{minipage}{0.35\textwidth}
\caption{Comparison of the ground and excited $0^+$ states calculated 
in the different basis sets: (i) the functions of Group I, 
(ii) the functions of Groups I and II, and (iii) the Pauli-allowed 
states obtained in the harmonic-oscillator basis. $E$ is the 
energy with respect to the 3$\alpha$ threshold. $\langle T \rangle$, 
$\langle V_{N} \rangle$, and $\langle V_{C} \rangle$ denote the 
expectation values of the kinetic energy, the nuclear potential energy, 
and the Coulomb potential energy, respectively. The $R_{\rm rms}$ 
value of Ref.~\cite{matsumura} is based on the $\alpha$ boson wave 
function which is obtained from a mapping of the microscopic 
3$\alpha$ wave function. All energies are in MeV 
and the rms radius is in fm.\vspace{12.2cm}}\label{comp.0+}
\end{minipage}~%
\quad
\begin{minipage}{0.4\textwidth}
\begin{tabular}{cccccccc}
\hline\hline
Basis & $Q_{\rm max}$ & State & $E$ & $\langle T \rangle$ & $\langle V_{N}
 \rangle$ & $\langle V_{C} \rangle$ & $R_{\rm rms}$ \\
\hline
  &   & $0^+_1$ & 0.40 & 74.34 & $-80.18$ & 6.24 & 2.03  \\
set (i)  &30 & $0^+_2$ & 7.27 & 61.91 & $-60.21$ & 5.56 & 2.28  \\
  &   & $0^+_3$ & 12.38 & 84.56 & $-78.19$ & 6.02 & 2.25  \\
\cline{2-8}
  &   & $0^+_1$ & $-0.02$ & 65.88 & $-71.84$ & 5.94 & 2.16  \\
  &40 & $0^+_2$ & 5.55 & 53.36 & $-53.04$ & 5.22 & 2.53  \\
  &   & $0^+_3$ & 8.89 & 61.94 & $-58.34$ & 5.29 & 2.56  \\
\cline{2-8}
  &   & $0^+_1$ &$-0.16$ & 61.77 & $-67.71$ & 5.78 & 2.25  \\
  &50 & $0^+_2$ & 4.53 & 48.11 & $-48.53$ & 4.96 & 2.76  \\
  &   & $0^+_3$ & 7.29 & 47.84 & $-45.33$ & 4.78 & 2.80  \\
\cline{2-8}
  &   & $0^+_1$ &$-0.22$ & 59.57 & $-65.48$ & 5.69 & 2.31  \\
  &60 & $0^+_2$ & 3.85 & 43.49 & $-44.34$ & 4.70 & 2.99  \\
  &   & $0^+_3$ & 6.44 & 39.45 & $-37.46$ & 4.46 & 3.00  \\
\hline
  &   & $0^+_1$ & $-19.90$ & 126.69 & $-155.08$ & 8.50 & 1.31  \\
set (ii)   &30 & $0^+_2$ & 0.26 & 69.44 & $-75.28$ & 6.10 & 2.06  \\
  &   & $0^+_3$ & 7.58 & 64.61 & $-62.63$ & 5.60 & 2.31  \\
\cline{2-8}
  &   & $0^+_1$ & $-19.90$ & 126.69 & $-155.08$ & 8.50 & 1.31  \\
  &40& $0^+_2$ &$-0.19$ & 61.44 & $-67.43$ & 5.81 & 2.20  \\
  &   & $0^+_3$ & 5.54 & 54.14 & $-53.81$ & 5.21 & 2.57  \\
\cline{2-8}
  &   & $0^+_1$ & $-19.90$ & 126.68 & $-155.08$ & 8.50 & 1.31  \\
  &50 & $0^+_2$ &$-0.34$ & 57.60 & $-63.59$ & 5.65 & 2.28  \\
  &   & $0^+_3$ & 4.48 & 48.49 & $-48.97$ & 4.96 & 2.77  \\
\cline{2-8}
  &   & $0^+_1$ & $-19.90$ & 126.68 & $-155.08$ & 8.50 & 1.31  \\
  &60 & $0^+_2$ &$-0.40$ & 55.60 & $-61.57$ & 5.57 & 2.34  \\
  &   & $0^+_3$ & 3.82 & 43.77 & $-44.67$ & 4.72 & 2.98  \\
\hline
  &   & $0^+_1$ & $-19.25$ & 128.33 & $-156.10$ & 8.53 & 1.31  \\
set (iii)  &30 & $0^+_2$ & $-0.11$& 63.82 & $-69.88$ & 5.95 & 2.09  \\
  &   & $0^+_3$ & 7.03 & 60.11 & $-58.57$ & 5.49 & 2.34  \\
\cline{2-8}
  &   & $0^+_1$ & $-19.25$ & 128.32 & $-156.09$ & 8.53 & 1.31  \\
  &40 &$0^+_2$ & $-0.59$& 55.55 & $-61.78$ & 5.64 & 2.23  \\
  &   & $0^+_3$ & 5.02 & 50.22 & $-50.34$ & 5.13 & 2.57  \\
\cline{2-8}
  &   & $0^+_1$ & $-19.25$ & 128.32 & $-156.09$ & 8.53 & 1.31  \\
  &50 & $0^+_2$ & $-0.75$& 51.75 & $-57.98$ & 5.49 & 2.32  \\
  &   & $0^+_3$ & 4.04 & 45.28 & $-46.15$ & 4.91 & 2.76  \\
\cline{2-8}
  &   & $0^+_1$ & $-19.25$ & 128.32 & $-156.09$ & 8.53 & 1.31  \\
  &60 & $0^+_2$ & $-0.81$& 49.86 & $-56.07$ & 5.41 & 2.36  \\
  &   & $0^+_3$ & 3.44 & 41.15 & $-42.40$ & 4.69 & 2.88  \\
\hline
Microscopic  &  & $0^+_1$ & $-7.72$ &  &  &  & 1.92  \\
(Ref.~\cite{matsumura})  & & $0^+_2$ & $0.71$&  &  &  & 4.26  \\
\hline
\end{tabular}
\end{minipage}~%
\end{center}
\end{table}

As demonstrated above, 
the two states of Group II play a key role in determining the 
characteristics of the macroscopic 3$\alpha$ model.  As shown in 
Table~\ref{APAS}, they contain a significant amount of 
the Pauli-allowed states with $Q(\lambda \mu)
=8(04), \,10(62), \,12(12,0), \,14(14,0)$ etc., which should be 
included from the microscopic point of view. The Hamiltonian 
matrix elements between these two states are given, in units of MeV,
as follows:
\begin{equation}
\begin{pmatrix}
14.58 & -7.07 \\
-7.07 & -7.78 
\end{pmatrix}.
\end{equation}
We see that the energy of the state with the 
eigenvalue $\gamma_2$ is already lower than the ground 
state energy obtained in the calculation with set (i). This 
is due to the fact that it is dominated by the lowest 
shell-model configuration of $Q(\lambda \mu)\!=\!8(04)$. 
The energy obtained by coupling the two states of Group II is $-9.8$ MeV, so they clearly 
contribute to lowering the energy. Whether these two states are included in the calculation or not, particularly the $\gamma_2$ state, is a crucial factor for producing quite different results.

\begin{figure}[t]
\begin{center}
\rotatebox{-90}{\includegraphics[width=8cm,clip]{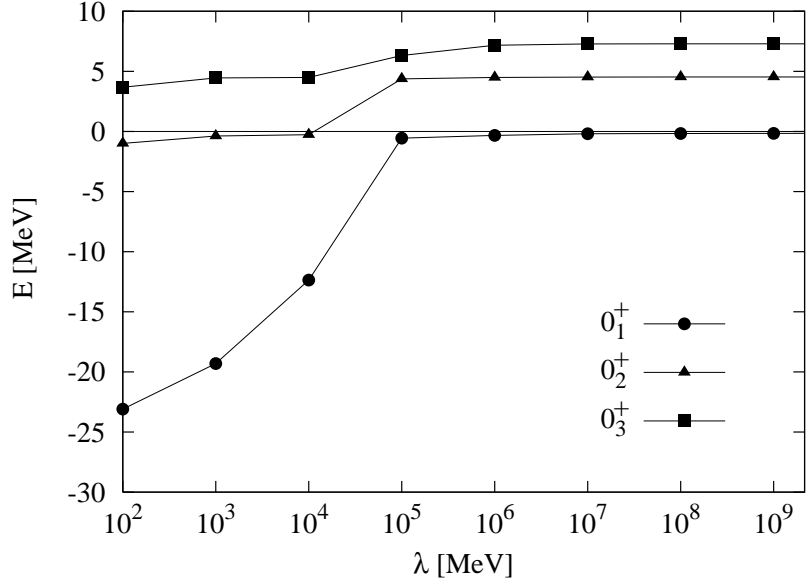}}
\end{center}
\caption{Convergence of the energy with respect to the 
3$\alpha$ threshold as a function of the 
projection constant $\lambda$. $Q_{\rm max}=50$. }
\label{e.vs.lambda}
\end{figure}

Next we discuss the origin of the enigmatic behavior of the 
energy $E(\lambda)$ which is obtained in the projection technique. 
We plot in fig.~\ref{e.vs.lambda} the energies of the ground 
and excited $0^+$ states as a function of the projection constant 
$\lambda$ in the case of $Q_{\rm max}=50$. 
Other choice of $Q_{\rm max}$ gives a similar result. 
The dependence on $\lambda$ follows the 
pattern observed in Refs.~\cite{tursunov,descouvemont,tursunov2} 
which use the basis functions different from the harmonic-oscillator 
basis. No indication of the 
energy convergence is attained for $\lambda < 10^5$ MeV, but it begins 
to show a convergence for larger $\lambda$ values. The origin of this 
behavior is attributed to the particular selection procedure 
of the states in the projection technique. It is clear that 
the states of Group III play no active role from the beginning 
because the eigenvalue $\gamma$ is of order unity and when it 
is multiplied by a large $\lambda$ value  
the energy expectation value becomes very high. 
On the contrary, the states of Group I are always active in the 
diagonalization of the pseudo Hamiltonian $\widetilde{H}$ as their 
eigenvalues are zero. The two states of Group II, 
especially the state with the eigenvalue 
$\gamma_2\!=\!1.07153\times 10^{-3}$, are in a subtle situation. 
The expectation value of $\widetilde{H}$ for this state is 
\begin{equation}
-7.78+\lambda\gamma_2,
\end{equation}
so the state is expected to contribute to the ground state 
significantly as long as $\lambda\gamma_2$ is smaller than, say 8 MeV, 
that is, for $\lambda < 10^4$ MeV, but it will be excluded in the end for sufficiently 
large $\lambda$. This is confirmed in Table~\ref{prob.als} which  
displays the probability of finding the state with $\gamma_2$ 
in the wave functions of the three 0$^+$ states. The probability 
found in the ground state is 0.64 for $\lambda\!=\!10^4$ MeV 
and it decreases rapidly for $\lambda \ge 10^5$ MeV.

\begin{table}[b]
\caption{$\lambda$ dependence of the 
probability of finding the state with the eigenvalue $\gamma_2$ 
in the ground and excited $0^+$ states in $^{12}$C. The 
calculations are performed for $Q_{\rm max}=50$. }
\label{prob.als}
\begin{center}
\begin{tabular}{ccccccccccccccc}
\hline\hline
& & \multicolumn{13}{c}{$\lambda$ (MeV)}\\
\cline{3-15} 
$^{12}$C & & $10^2$ & & $10^3$ & & $10^4$ & & $10^5$ & & $10^6$ & & $10^7$ & & $10^8$ \\
\hline
$0^+_1$ & & 7.16$\times 10^{-1}$ & & 7.46$\times 10^{-1}$ & & 6.44$\times 10^{-1}$ & & 9.62$\times 10^{-4}$ & & 1.74$\times 10^{-6}$ & & 3.03$\times 10^{-9}$ & & 1.76$\times 10^{-11}$ \\
 $0^+_2$ & & 1.87$\times 10^{-2}$ & & 6.05$\times 10^{-3}$ & & 1.04$\times 10^{-2}$ & & 1.07$\times 10^{-3}$ & & 6.20$\times 10^{-7}$ & & 9.76$\times 10^{-10}$ & & 6.17$\times 10^{-12}$ \\
$0^+_3$ & & 4.83$\times 10^{-3}$ & & 1.07$\times 10^{-3}$ & & 1.64$\times 10^{-3}$ & & 9.83$\times 10^{-3}$ & & 1.67$\times 10^{-5}$ & & 1.04$\times 10^{-7}$ & & 9.81$\times 10^{-10}$ \\
\hline
\end{tabular}
\end{center}
\end{table}

\section{$\alpha\!-\!\alpha$ potential}
\label{alpha-alphapot}

In the previous section we have argued that the two-body forbidden 
states for the $\alpha\!-\!\alpha$ relative motion 
have to be defined 
referring to the microscopic wave function of the 
$\alpha$ particle, not from the bound states of the 
$\alpha\!-\!\alpha$ potential. Based on this viewpoint, we have 
shown that the BFW potential is so attractive that the 
ground state energy of ~$^{12}$C is strongly bound. Here we 
discuss an $\alpha\!-\!\alpha$ potential which is more 
suitable for the macroscopic calculation. In doing so, we 
impose the condition that the $\alpha\!-\!\alpha$ relative 
motion should be 
orthogonal to the 2$\alpha$ Pauli-forbidden states, that is, the 
harmonic-oscillator functions $\psi^q_{\ell m}$ with $q=0,2$.

First we modify the strength of the BFW potential so as to 
reproduce the $0^+$ resonance energy of the 2$\alpha$ system 
under the above orthogonality constraint. The resulting potential, 
which we call MBFW, has a strength parameter of 
$v_0=-121.888$ MeV. Note, however, that this modified potential 
does not reproduce the 2$^+$ resonance energy. 
Table~\ref{energy.new} lists the energies and $R_{\rm rms}$ values 
calculated with this potential in the basis functions of set (iii) 
with $Q_{\rm max}\!=\!50$. They are almost the same as those 
obtained with the BFW potential. Compare these with 
the result of set (iii) in Table~\ref{comp.0+}. Since no 
improvement is 
attained by the MBFW potential, we conclude that the change of $v_0$ 
of the BFW potential does not lead to an $\alpha\!-\!\alpha$ 
potential suitable for the macroscopic calculation. 

It seems that the range of the BFW potential has no direct 
connection to the underlying nucleon-nucleon potential or to the wave 
function of the $\alpha$ particle. Another potential we consider 
here is similar to the $\alpha\!-\!\alpha$ folding potential 
used in Ref.~\cite{kurokawa}. We assume 
the same form as the BFW potential, and choose the potential 
parameters as follows:
\begin{equation}
v_0=-107.9 \ {\rm MeV},\ \ \ \rho_0=0.20 \ {\rm fm}^{-2},
\ \ \ \beta=0.589 \ {\rm fm}^{-1}.
\label{mos}
\end{equation}
Again, this potential reproduces the $0^+$ resonance of $^8$Be, 
but predicts the $2^+$ resonance to be about 1.9 MeV, which appears 
too low compared to the experimental value, $E\!=\!3.13$ MeV. 
The range of this potential is longer than that of the BFW 
potential, and consequently its strength is made considerably 
weaker. The energies and 
$R_{\rm rms}$ values of the ground and excited $0^+$ states 
are obtained with this new potential in the Pauli-allowed space of 
set (iii) and they are listed in Table~\ref{energy.new}. 
The energy of the ground state is now pretty much improved, 
though it is still too strongly bound by about 4 MeV. 

\begin{table}[t]
\caption{Comparison of the ground and excited $0^+$ states 
calculated 
with the different $\alpha\!-\!\alpha$ potentials. The basis 
states are either harmonic-oscillator functions of set (iii) 
with $Q_{\rm max}=50$, or the $\alpha$ boson wave functions  
defined in Eq.~(\ref{alpha.boson}). See the caption of 
Table~\ref{comp.0+}. }
\label{energy.new}
\begin{center}
\begin{tabular}{cccccccc}
\hline\hline
Potential & Basis & State & $E$ & $\langle T \rangle$ & $\langle V_{N}
 \rangle$ & $\langle V_{C} \rangle$ & $R_{\rm rms}$ \\
\hline
  &   & $0^+_1$ & $-18.32$ & 126.58 & $-153.37$ & 8.48 & 1.32  \\
MBFW & set (iii) & $0^+_2$ & $-0.41$ & 49.44 & $-55.25$ & 5.40 & 2.36  \\
  &   & $0^+_3$ & 4.31 & 44.92 & $-45.50$ & 4.90 & 2.76  \\
\hline
  &   & $0^+_1$ & $-11.68$ & 95.66 & $-114.33$ & 6.99 & 1.52  \\
Eq.~(\ref{mos})&set (iii)& $0^+_2$ & 0.63 & 35.18 & $-39.18$ & 4.63 & 2.61  \\
  &   & $0^+_3$ & 5.00 & 37.94 & $-37.46$ & 4.52 & 2.79  \\
\hline
Eq.~(\ref{mos})&$\alpha$-boson   & $0^+_1$ & $-11.69$ & 95.87 & $-114.56$ & 6.99 & 1.52  \\
& & $0^+_2$ & $0.21$& 25.98 & $-29.92$ & 4.15 & 3.17  \\
\hline
\end{tabular}
\end{center}
\end{table}

The truncation with $Q_{\rm max}\!=\!50$ may not be good enough 
to obtain convergent results for the excited states. 
To check this, we diagonalize the 
Hamiltonian in $\alpha$-boson functions $g$ which are obtained 
from correlated Gaussians $G$ through the following fermion$\to$boson 
mapping: 
\begin{equation}
g(a,{\vi r}_1,{\vi \rho}_1)
=\langle \phi(\alpha_1)\phi(\alpha_2)\phi(\alpha_3)\mid 
{\cal A}\big\{\phi(\alpha_1)\phi(\alpha_2)\phi(\alpha_3)
G(a,{\vi R})\big\} \rangle,
\label{alpha.boson}
\end{equation}
with
\begin{equation}
G(a,{\vi R})={\rm exp}\Big\{-a_{12}({\vi R}_1-{\vi R}_2)^2 
-a_{13}({\vi R}_1-{\vi R}_3)^2-a_{23}({\vi R}_2-{\vi R}_3)^2
\Big \},
\end{equation}
where $\phi(\alpha)$ is the intrinsic wave function of the 
$\alpha$ particle, and ${\cal A}$ is an operator which 
antisymmetrizes the 3$\alpha$ microscopic wave function 
built on the 12-nucleon coordinates. A parameter 
$a=(a_{12},a_{13},a_{23})$ specifies the 
correlated Gaussian. By choosing the parameter $a$ 
appropriately, one can express a variety of different 
shapes of the three-body system. The function $g$ obviously has [3] 
symmetry, i.e., 
$g(a,{\vi r}_1,{\vi \rho}_1)\!=\!g(a,{\vi r}_2,{\vi \rho}_2)
\!=\!g(a,{\vi r}_3,{\vi \rho}_3)$ and 
$g(a,-{\vi r}_1,{\vi \rho}_1)\!=\!g(a,{\vi r}_1,{\vi \rho}_2)$ etc. 
Furthermore, it is orthogonal to 
the Pauli-forbidden states 
$\psi^q_{\ell m}$ ($q=0,2$) if $\phi(\alpha)$ is constructed from 
the $(0s)^4$ harmonic-oscillator function (with its center of mass 
part being eliminated). Thus the function 
$g$, for arbitrary $a$, satisfies all the conditions necessary for 
the allowed states as discussed in sect.~\ref{ashosm}. 
Moreover, it should be noted that the boson function $g$ contains 
not only the two-body Pauli effects but also full 
three-body exchange effects. The method of calculating $g$ is 
given in Ref.~\cite{matsumura}. 

The 3$\alpha$ wave function $\Psi$ is now approximated with  
a combination of $g$. The parameters of $g$ are selected by 
the stochastic variational method~\cite{book}. The results of 
this calculation are also given in Table~\ref{energy.new}. 
It is seen that the harmonic-oscillator expansion converges 
very slowly except for the ground state. The $R_{\rm rms}$ value 
for the $0^+_2$ state increases to 3.17 fm, which is still smaller 
than the value calculated in the microscopic model. 
As noted above, the calculation using $g$ includes not only 
the two-body Pauli effects but also the three-body exchange 
effects, whereas the calculation of set (iii) includes only  
the two-body Pauli effects. The fact that they give virtually 
the same result for the ground state indicates that the 
three-body exchange effects are negligibly small. Thus we 
do not agree with the discussion of Ref.~\cite{tursunov} 
,which was made for a possible reason of the sensitivity of 
the projection technique.

Though a considerable improvement is obtained with the new 
potential, the ground state energy is still too low by about 
4 MeV. One possible reason for this is that the potential is 
too attractive for the $d$ wave 
(and probably for the $g$ wave as well). Then the lowest Pauli-allowed 
state of $Q(\lambda \mu)\!=\!8(04)$ with $L\!=\!0$, for example, 
gains too much energy, because the dominating partial waves of 
this state 
are $d$ and $g$ waves, as understood from the SU(3) coefficients  
$\langle (40)\ell, (40)\ell \mid\mid(04) 0\rangle^2$, which are 
0.28, 0.36, and 0.36 for $\ell=0,2$, and 4, respectively. 

The above consideration suggests that a more suitable 
$\alpha\!-\!\alpha$ potential should have an $\ell$-dependence. 
In this respect, it is interesting to recall 
a series of recent publications~\cite{fujiwara1,fujiwara2,fujiwara5}, 
where the utility of a 2$\alpha$ resonating group method (RGM) kernel 
has been investigated for the 3$\alpha$ system. The RGM kernel is 
non-local and energy-dependent, and therefore a solution of the system 
with such a potential requires some sort of 
self consistency procedure. The ground state energy of $^{12}$C 
turns out to be quite acceptable, that is, it is not strongly bound 
and is predicted to be around the value obtained by the microscopic 
calculation. Since a non-local potential could be converted to an 
equivalent local potential with $\ell$ dependence, an 
$\alpha\!-\!\alpha$ potential suitable for a macroscopic calculation 
may be found in $\ell$-dependent form.

\section{Summary}
\label{summary}
We studied the ground and excited 0$^+$ states of $^{12}$C 
in the 3$\alpha$ model, using the deep BFW 
potential. The three redundant states with $0s,\, 1s$, and $0d$ 
of this potential were eliminated from the function space of 
the 3$\alpha$ system. Two methods of elimination were tested. One is to diagonalize $\Gamma$ in 
a complete set of translationally invariant 
harmonic-oscillator functions with [3] symmetry in order to 
separate the allowed states from the forbidden ones. The other 
is to use the orthogonalizing pseudopotential $\lambda \Gamma$ and 
to obtain the energy which converges for 
$\lambda$ large enough to kill the 
contribution of the forbidden states. These methods, though 
equivalent, were tested in order to clarify not only 
the origin of the enigmatic $\lambda$-behavior of the second 
method, but also to give a reason why the ground state energy is 
very high. 

We found two eigenstates of $\Gamma$ with almost zero 
eigenvalues. One of them especially plays a key role. This one has a dominant configuration corresponding to the lowest shell-model state of 
$Q(\lambda \mu)=8(04)$, and in spite of its small eigenvalue, of order 10$^{-3}$, it was eliminated as a forbidden state in the 
orthogonalizing pseudopotential approach. The energy expectation 
value of this state alone is already close to the ground state 
of $^{12}$C. With the exclusion of this state, there is no hope to get a ground state energy close to the experiment. 
In addition, we showed that the peculiar dependence of the 
energy on $\lambda$ 
can be understood from the admixture of this state in the 
solution: for 
$\lambda < 10^{4}$ MeV, this state can have the effect of lowering the energy, but for $\lambda > 10^5$ MeV it can no longer keep this effect in the ground state energy. 

The existence of the states with almost zero eigenvalue of $\Gamma$ 
is not a characteristic of the BFW potential, but also applies to 
other $\alpha\!-\!\alpha$ potentials provided their bound states are well approximated with the three 
harmonic-oscillator functions $\psi^q_{\ell m}\ (q=0,2)$. We find that 
the potential of Eq.~(\ref{mos}) produces two almost Pauli-allowed 
states with $\gamma_1=1.59640\times 10^{-5}$ and 
$\gamma_2=1.17812\times 10^{-3}$ for $Q_\mathrm{max}=50$. 

In our opinion the eigenstates with almost zero eigenvalue should 
be included in the 3$\alpha$ allowed space as they correspond to 
the most important shell-model configurations, otherwise, the 
3$\alpha$ macroscopic model loses a link to the microscopic 
3$\alpha$ model. To include these states automatically, it is 
safe to define the operator $\Gamma$ in terms of the 
harmonic-oscillator functions. 
With this definition of $\Gamma$ we demonstrated that 
the BFW potential is too attractive and an appropriate $\alpha\!-\!\alpha$ potential has to be 
determined referring to the underlying microscopic model.


\begin{acknowledgments}
This work was in part supported by a Grant for Promotion of 
Niigata University Research Project (2005--2007).
\end{acknowledgments}


\section*{\large{Appendix: Pauli-allowed states in SU(3) coupled basis}}
We give the Pauli-allowed states, 
$\chi_{00}^{[3]Q(\lambda \mu)\kappa \gamma=0}$, 
for $Q=8-14$ in terms of the SU(3) coupled basis. 
The index $\kappa$ is suppressed when it is unnecessary.

\begin{itemize}
\item $Q=8$
\begin{eqnarray}
&&\chi_{00}^{[3]8(04)\gamma=0}=[\psi^4({\vi r}_1)\psi^4({\vi
 \rho}_1)]^{(04)}_{00}.
\end{eqnarray}
\item $Q=10$
\begin{eqnarray}
&&\chi_{00}^{[3]10(24) \gamma=0}
=\frac{1}{\sqrt{2}}[\psi^4({\vi r}_1)\psi^6({\vi \rho}_1)]^{(24)}_{00}
+\frac{1}{\sqrt{2}}[\psi^6({\vi r}_1)\psi^4({\vi \rho}_1)]^{(24)}_{00},
\nonumber \\
&&\chi_{00}^{[3]10(62)\gamma=0}
=\frac{3\sqrt{3}}{2\sqrt{11}}[\psi^4({\vi r}_1)\psi^6({\vi \rho}_1)]^{(62)}_{00}
-\frac{\sqrt{3}}{\sqrt{11}}[\psi^6({\vi r}_1)\psi^4({\vi \rho}_1)]^{(62)}_{00}
\nonumber \\
&&\qquad\qquad\quad
 +\, \frac{\sqrt{5}}{2\sqrt{11}}[\psi^8({\vi r}_1)\psi^2({\vi
 \rho}_1)]^{(62)}_{00}.
\end{eqnarray}
\item $Q=12$
\begin{eqnarray}
&&\chi_{00}^{[3]12(06) \gamma=0}
=[\psi^6({\vi r}_1)\psi^6({\vi \rho}_1)]^{(06)}_{00},
\nonumber \\
&&\chi_{00}^{[3]12(44)\gamma=0}
=\frac{\sqrt{3}}{2\sqrt{2}}[\psi^4({\vi r}_1)\psi^8({\vi \rho}_1)]^{(44)}_{00}
+\frac{1}{2}[\psi^6({\vi r}_1)\psi^6({\vi \rho}_1)]^{(44)}_{00}
\nonumber \\
&&\qquad\qquad\quad 
+\, \frac{\sqrt{3}}{2\sqrt{2}}[\psi^8({\vi r}_1)\psi^4({\vi
 \rho}_1)]^{(44)}_{00},
\nonumber \\
&&\chi_{00}^{[3]12(82)\gamma=0}
=\frac{9\sqrt{5}}{4\sqrt{43}}[\psi^4({\vi r}_1)\psi^8({\vi \rho}_1)]^{(82)}_{00}
+\frac{3}{2\sqrt{86}}[\psi^6({\vi r}_1)\psi^6({\vi \rho}_1)]^{(82)}_{00}
\nonumber \\
&&\qquad\qquad\quad
-\, \frac{5\sqrt{5}}{4\sqrt{43}}[\psi^8({\vi r}_1)\psi^4({\vi
 \rho}_1)]^{(82)}_{00}+\frac{\sqrt{35}}{2\sqrt{43}}[\psi^{10}({\vi
 r}_1)\psi^2({\vi \rho}_1)]^{(82)}_{00},
\nonumber \\
&&\chi_{00}^{[3]12(12,0)\gamma=0}
=\frac{27\sqrt{7}}{116}[\psi^4({\vi r}_1)\psi^8({\vi \rho}_1)]^{(12,0)}_{00}
-\frac{9\sqrt{15}}{58}[\psi^6({\vi r}_1)\psi^6({\vi \rho}_1)]^{(12,0)}_{00}
\nonumber \\
&&\qquad\qquad\quad\;\;
+\, \frac{9\sqrt{7}}{58}[\psi^8({\vi r}_1)\psi^4({\vi
 \rho}_1)]^{(12,0)}_{00}-\frac{\sqrt{105}}{29\sqrt{2}}[\psi^{10}({\vi
 r}_1)\psi^2({\vi \rho}_1)]^{(12,0)}_{00}
\nonumber \\
&&\qquad\qquad\quad\;\;
+\,\frac{\sqrt{385}}{116}[\psi^{12}({\vi
 r}_1)\psi^0({\vi \rho}_1)]^{(12,0)}_{00}.
\end{eqnarray}
\end{itemize}

We have two Pauli-allowed states for $Q(\lambda \mu)=14(64)$, 
so there is a freedom to fix them. Here they 
are defined in such a way that one of them $(\kappa=2)$ contains no 
$[\psi^{10}({\vi r}_1)\psi^{4}({\vi \rho}_1)]^{(64)}_{00}$ component.

\newpage 

\begin{itemize}
\item $Q=14$
\begin{eqnarray}
&&\chi_{00}^{[3]14(26) \gamma=0}
=\frac{1}{\sqrt{2}}[\psi^6({\vi r}_1)\psi^8({\vi \rho}_1)]^{(26)}_{00}
+\frac{1}{\sqrt{2}}[\psi^8({\vi r}_1)\psi^6({\vi \rho}_1)]^{(26)}_{00},
\nonumber \\
&&\chi_{00}^{[3]14(64)\kappa=1 \gamma=0}
=0.479702[\psi^4({\vi r}_1)\psi^{10}({\vi \rho}_1)]^{(64)}_{00}
+0.619292[\psi^6({\vi r}_1)\psi^8({\vi \rho}_1)]^{(64)}_{00}
\nonumber \\
&&\qquad\qquad\qquad\;\,
+\, 0.206431[\psi^8({\vi r}_1)\psi^{6}({\vi \rho}_1)]^{(64)}_{00}
+0.586302[\psi^{10}({\vi r}_1)\psi^{4}({\vi \rho}_1)]^{(64)}_{00},
\nonumber \\
&&\chi_{00}^{[3]14(64)\kappa=2 \gamma=0}
=0.337100[\psi^4({\vi r}_1)\psi^{10}({\vi \rho}_1)]^{(64)}_{00}
-0.522233[\psi^6({\vi r}_1)\psi^8({\vi \rho}_1)]^{(64)}_{00}
\nonumber \\
&&\qquad\qquad\qquad\;\,
+\, 0.783349[\psi^8({\vi r}_1)\psi^6({\vi \rho}_1)]^{(64)}_{00},
\nonumber \\
&&\chi_{00}^{[3]14(10,2)\gamma=0}
=\frac{9\sqrt{7}}{8\sqrt{19}}[\psi^4({\vi r}_1)\psi^{10}({\vi \rho}_1)]^{(10,2)}_{00}
+\frac{3\sqrt{3}}{2\sqrt{38}}[\psi^6({\vi r}_1)\psi^8({\vi \rho}_1)]^{(10,2)}_{00}
\nonumber \\
&&\qquad\qquad\quad\;\,\,
-\, \frac{\sqrt{3}}{4\sqrt{38}}[\psi^8({\vi r}_1)\psi^6({\vi
 \rho}_1)]^{(10,2)}_{00}
-\frac{\sqrt{7}}{2\sqrt{19}}[\psi^{10}({\vi r}_1)\psi^4({\vi
 \rho}_1)]^{(10,2)}_{00}
\nonumber \\
&&\qquad\qquad\quad\;\,\,
 -\,\frac{3\sqrt{35}}{8\sqrt{19}}[\psi^{12}({\vi r}_1)\psi^2({\vi
 \rho}_1)]^{(10,2)}_{00},
\nonumber \\
&&\chi_{00}^{[3]14(14,0)\gamma=0}
=\frac{81}{4\sqrt{742}}[\psi^4({\vi r}_1)\psi^{10}({\vi \rho}_1)]^{(14,0)}_{00}
-\frac{9\sqrt{3}}{4\sqrt{742}}[\psi^6({\vi r}_1)\psi^8({\vi \rho}_1)]^{(14,0)}_{00}
\nonumber \\
&&\qquad\qquad\quad\;\,\,
-\, \frac{9\sqrt{3}}{2\sqrt{742}}[\psi^8({\vi r}_1)\psi^6({\vi
 \rho}_1)]^{(14,0)}_{00}+\frac{21}{2\sqrt{742}}[\psi^{10}({\vi
 r}_1)\psi^4({\vi \rho}_1)]^{(14,0)}_{00}
\nonumber \\
&&\qquad\qquad\quad\;\,\,
-\,\frac{11\sqrt{11}}{4\sqrt{742}}[\psi^{12}({\vi r}_1)\psi^2({\vi \rho}_1)]^{(14,0)}_{00}
+\frac{\sqrt{143}}{4\sqrt{106}}[\psi^{14}({\vi r}_1)\psi^0({\vi
\rho}_1)]^{(14,0)}_{00}.
\nonumber \\
\end{eqnarray}
\end{itemize}

\end{document}